\documentclass[preprint,12pt,3p]{elsarticle}

 \usepackage{graphicx}
 \usepackage{subcaption}
\usepackage{verbatim}
\usepackage{amssymb}
\usepackage{float}
\usepackage{gensymb}
\usepackage[table,xcdraw]{xcolor}
\newcommand{\quotes}[1]{``#1''}
\usepackage[none]{hyphenat}

\journal{Elsevier}

\begin{document}

\begin{frontmatter}

\title{Numerical Analysis of Sensitivity Enhancement of Surface Plasmon Resonance Biosensors Using a Mirrored Bilayer Structure}

\author{Sayeed Shafayet Chowdhury,$^{1,*}$ and Syed Mohammad Ashab Uddin,$^{2}$}
%\address{$^1$Bangladesh University of Engineering and Technology, Dhaka 1205, Bangladesh. \\$^+$These authors contributed equally to this work.}
\address{$^1$Purdue University, West Lafayette, IN 47907, USA,}
\address{$^2$University of California,  Irvine, CA 92697, USA. }
\address{$^*$Correspondence Email: chowdh23@purdue.edu}

\begin{abstract}
A mirrored bilayer structure incorporating layers of Graphene and MoS$_2$ is proposed here for Surface Plasmon Resonance (SPR) biosensing and its performance is evaluated numerically. Starting from the basic configuration,
the structure with graphene and MoS$_2$ layers is  gradually developed for enhanced performance. Reflectance is the main considered parameter for performance analysis.
A theoretical framework based
on Fresnel$'$s equations is presented and by measuring reflectance versus angle of incidence, sensitivity is calculated from the displacement of SPR
angle using finite-difference time-domain (FDTD) technique. Our numerical analysis shows that using the proposed approach, 
about 4.2 times enhanced sensitivity can be achieved compared to the basic Kretschmann
configuration. Notably, the structure provides enhancement for both angular and wavelength interrogations. Furthermore, simulations have been repeated for various ligate-ligand pairs and consistent enhancement has been observed which proves the robustness
of the structure. So, the proposed sensor architecture clearly provides pronounced improvement in sensitivity which can be significant in various practical applications.
\end{abstract}


\end{frontmatter}

% \linenumbers

\section{Introduction}
\label{sec1}

In recent times, there has been immense growth in usage of optical biosensors in disease detection due to highly specific, cost effective and sensitive applications \cite{kurihara, ppg1, ppg2}. Plasmonics is an attractive research arena in this regard as it provides real-time and label-free detection of different biosubstances. Efforts focusing on sensitivity enhancement lies at the core of surface plasmon resonance (SPR) biosensing research as it paves the way for detecting tiny variations in the sensing layer which can be critical in revealing disease onsets. 

SPR spectroscopy is extremely suitable for analyzing biomolecular interactions occurring in the vicinity of sensor surfaces \cite{2, nanowire}. When a biomolecule on a sensor surface meets its partner in a solution, it creates a binding affinity interaction that in turn changes the interfacial refractive index observable by the attached sensor \cite{collagen, ecoli, our1}. Liedberg $et$ $al.$ \cite{liedberg} first demonstrated application of SPR in immunology. In an elaborate review, summary of SPR biosensors was presented by Homola \cite{homola}.
SPR is the excitation of Surface Plasmon Waves (SPWs) which was first introduced by Otto \cite{otto}. However, Kretschman successfully commercialized it through excitation of SPWs on a silver film adhered to glass substrate \cite{krets}. Later Nylander and Liedberg $et$ $al.$ demonstrated the suitability of the Kretschmann configuration based SPR for both gas and biomolecular sensing \cite{nylander}. Since then, a diverse range of sensing applications have been benefited from this technique \cite{suvarna, lertva}. Chiu $et$ $al.$ was the first to report the use of Graphene-Oxide based SPRs for analyzing sensitivity \cite{chiu}, which has an alternative model with Graphene multilayer. This multilayer can induce enhanced electric field by virtue of strong coupling at the interface when deposited on metallic thin film or functionalized by Gold or Silver nanoparticles, but the field intensity depends on number of layers used \cite{zeng}.
Recently, it has gained popularity to incorporate additional composites with superior optical properties like CNTs \cite{sun}, magnetic nanoparticles \cite{jia}, gold nanoparticles \cite{li}, nanoelectronics \cite{du}, and P-doped Si crystals for sensitivity improvement \cite{rowe}. Wang $et$ $al$. \cite{trilayer} and some other researchers have experienced enhancement employing more than one metal layer. Artar $et$ $al.$ have shown that Fabry–-Perot (FP) cavity resonances in a  multilayered  plasmonic  crystals is highly  sensitive  to  refractive index changes \cite{artar}. According to \cite{shuwen}, perovskite-based metasurfaces have also provided sensitivity improvement. Furthermore, ultra-sensitive sensing has been achieved through multilayered self-assembly of graphene oxide and its reductions \cite{chung}. However, most of these approaches require significant fabrication efforts with increased complexity. Hence, it is desirable to have performance enhancement through simple structural modifications to a basic SPR biosensor without altering the process flow greatly. 

To that effect, study on enhancement of  sensitivity of SPR based biosensor is presented in this article. Here, we use lipid as a sensing layer since lipid bilayers are the basic component of the cell membrane. For any disease to attack a healthy cell, its carrier has to penetrate the membrane first. As a result, changes in the membrane become an important point of study as molecular targets for validated drugs and disease linking, which in turn necessitates study of lipids. We have proposed an SPR based novel biosensor structure using Kretschmann configuration where the bond-forming region is mirrored through reflection over the channel, effectively doubling the possible volume of bond formation. Here, graphene plays role as
bio-recognition constituent through pi-stacking force, while MoS$_2$ layers enhance light absorption by assisting in efficient charge transfer. We use finite-difference time-domain (FDTD) simulations to validate our idea and show that our proposed bilayer structure increases the sensitivity over the conventional planar SPR sensor based on Kretchmann configuration by approximately $320\%$ for angular interrogation. Again, as the detection mechanism here is optical, it is mainly concerned with the refractive index variation of the sensing volume. As a result, the obtained sensitivity enhancement is largely independent of sensing layer composition.

\section{Development of the Proposed Structure}
In this section, we gradually develop our proposed structure for enhanced sensitivity for lipid detection. We start with the basic structure containing the ligand and sensing layers on top of the prism as showed in Fig.~1(a). According to \cite{our2}, we optimized the thickness of the metal layer. When a solution is flown through the channel, the phospholipid
molecules (the sensing element here) bind with the tryptophan molecules (ligands in this case). The detailed performance analysis of this structure is given in section 4. This approach cannot provide very high sensitivity. So, to increase the sensitivity, we adopt a bilayer approach.
\begin{figure}[H]
\centering
\graphicspath{ {Thesis/} }
\includegraphics[height=4in, width = 5in]{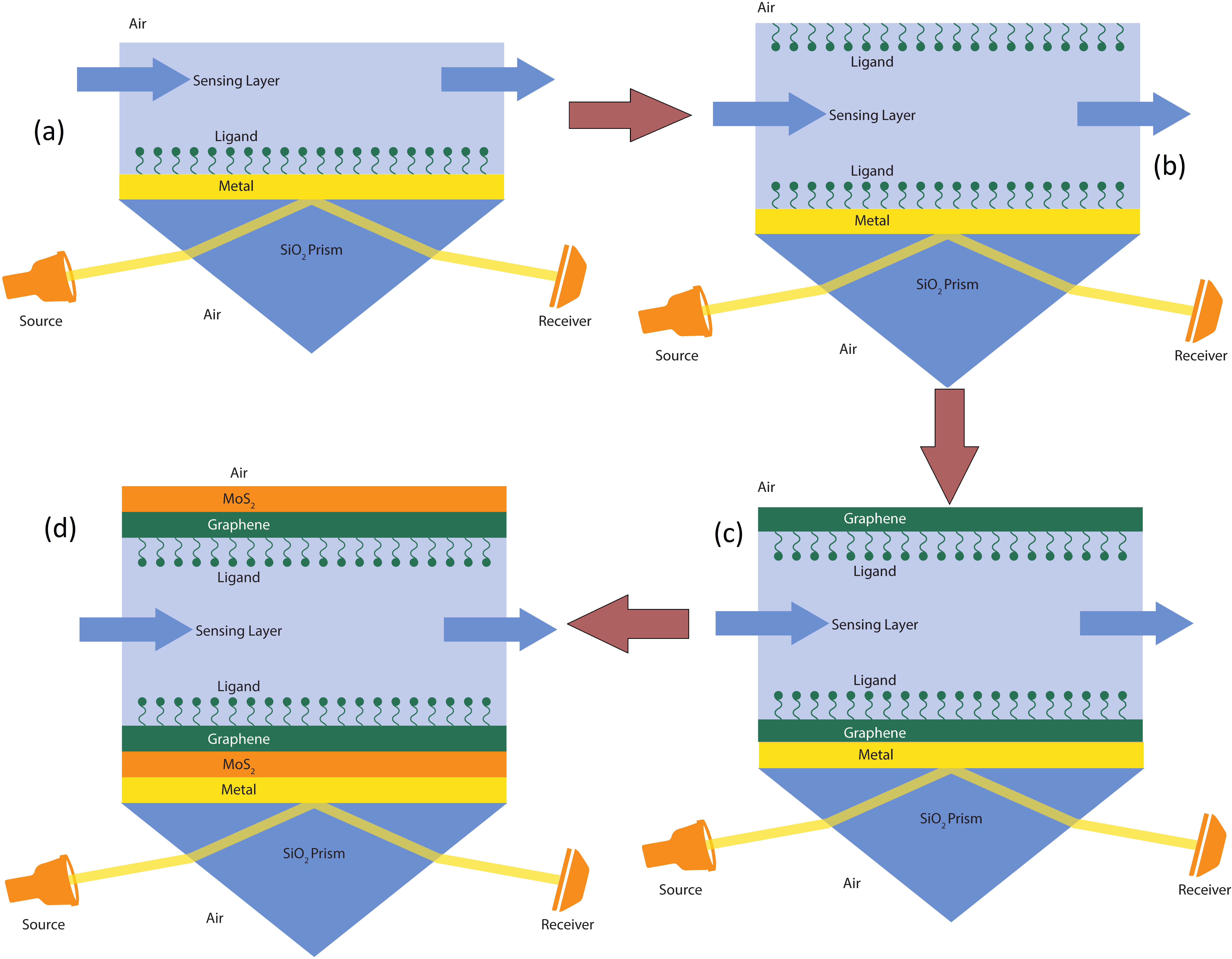} 
\caption{ Evolution of proposed structure, (a) basic structure, (b) bilayer SPR sensor configuration, (c) bilayer configuration with graphene and (d) proposed SPR sensor structure with mirrored layers of graphene and MoS$_2$.}
\end{figure}

In SPR sensors, the shift in SPR angle when the binding occurs is what  determines the sensitivity. The higher the shift, the more sensitive the sensor is. Now for the bilayer structure, Fig.~1(b) shows that the flow channel is sandwiched between two ligand layers. Here the sensing
molecules form bonds with the ligands on either side of the channel which results in a larger change in the effective refractive index of the sensing layer. This enhanced shift of refractive index eventually leads to a higher shift of SPR angle and hence, higher sensitivity.
To further enhance the sensitivity and improve performance, a layer of graphene on each side of the ligands is used as displayed in Fig.~1(c). In this respect, it is important to be able to anchor the ligands properly to the corresponding receptors. The graphene layer is useful in this regard as it aids in new surface functionalization. Additionally, graphene layers are highly hydrophilic and help to passivate the sensor surface against oxidation. Moreover, the ability to vary the graphene layer thickness acts as a tuning option for the designer to obtain the desired SPR behavior. %Because of the specific nature of graphene,  [29,45,46].
Large surface area and significant field enhancement are also among other benefits when graphene is used \cite{zeng}. %The researchers also determined that utilization of doped monolayer graphene can bring about more improvements in the applications currently available for fabrication of quantum plasmonic and atomic-scale nanoplasmonic instruments. 

Extremely thin layers of MoS$_2$ have been shown to amplify the sensor performance significantly \cite{maurya}. Monolayer MoS$_2$ provides a high optical absorption efficiency (around 5$\%$) and a direct bandgap of 1.8 eV \cite{zeng}. So, considering these salient features, we applied graphene with MoS$_2$ in the bilayer structure which forms our proposed structure. The proposed architecture is schematically depicted in Fig.~1(d). The first layer from the bottom is a glass prism, which changes the momentum of the incident light so that surface plasmons can be excited. A 50 nm gold layer is deposited on the prism, thickness of this metal layer has been optimized as per \cite{our2}. This is followed by a monolayer MoS$_2$ of 0.65 nm. Monolayer is chosen to have direct bandgap. Then there are three layers of graphene with a total thickness of 1.035 nm. Though at times sensitivity benefits from increasing depth of graphene layers , it also leads to broadening of the reflection curve, resulting in lower detection accuracy \cite{kumar}. In this study, three layers of graphene is chosen since it has been found to provide most optimized results. The next part of
the structure is for the binding and flow of biomolecular samples. The structure is a mirror image except for the glass prism and metal across the x-y plane through half-way of the channel. In case of detecting lipids, e.g. phospholipid using the proposed
structure, amino acid tryptophan are used as ligands. In this work, we assume a monolayer of tryptophan of 1.3 nm on graphene layer. The solution containing the phospholipid is flowed through one end of the channel and after detection, the phospholipid is washed away through
the other end. In this proposed sensor architecture, light absorption is enhanced by the MoS$_2$ layer through providing sufficient excitation energy for effective charge transfer, whereas, graphene aids in bio-recognition through pi-stacking force.

\section{Mathematical framework of the proposed structure}
To analyze the SPR phenomenon, TM-polarized light is transmitted from a light source towards a prism with large refractive index. As the incidence angle is gradually increased from a small value, total
internal reflection occurs after the critical angle is passed. An evanescent wave is generated subsequently which penetrates the metal as well as adjacent sensing layer. With further increase in angle of incidence, energy from this evanescent wave gets completely absorbed by the surface plasmons excited in the thin metal
film at the SPR angle ($\theta_{\textsubscript{SPR}}$). As a result, the reflected light strength becomes minimum in that case. The following formalism begins with the Kretschmann sensor layout and from there it is
extended towards the concept for our proposed structure. The sensing principle for the basic configuration can be described by \cite{kurihara}-
\begin{equation}
n_p\frac{\omega}{c}\sin(\theta_{SPR}) = \sqrt{\frac{\epsilon_m\epsilon_s}{\epsilon_m + \epsilon_s}}.
\end{equation}
Here, $n_p$ is a refractive index of prism; $\epsilon_m$ and $\epsilon_s$ are the permittivity of the metal and sensing layer. This can be approximated by the following expression-
\begin{equation}
n_psin\theta_{SPR} \equiv n_s,
\end{equation}
where, $n_s$ is the refractive index of sensing layer. For our proposed structure, we have multiple layers. According to \cite{kurihara}, if we have multiple layers after metal layer, we can use effective refractive index $n_{eff}$ for sensing layer and sensing equation can be approximated as 
\begin{equation}
sin\theta_{SPR} = \frac{n_{eff}}{n_p},
\end{equation}
where $n_{eff}$ is the effective refractive index of the the layers on the other
side of the prism. It is defined as
\begin{equation}
n_{eff} = \frac{2}{l_d}\int_{0}^{\infty} n(z)e^{-2z/l_d}dz,
\end{equation}
where $l_d$ is the penetration depth. Now for the proposed structure with multiple layers of dielectrics on the
upper part of the metal, we develop the necessary equations by extending the formalism in \cite{kurihara} for the reflections measurements as,
\begin{equation}
  R = \left |r_{0,9}\right|^2,  
\end{equation}
with,
\begin{equation}
  r_{i,9} = \frac{(r_{i,i+1}+r_{i+1,9}\times\exp(2ik_{zi+1}d_{i+1}))} {(1+r_{i,i+1}\times r_{i+1,9}\times\exp(2ik_{zi+1}d_{i+1}))}, 
\end{equation}
\begin{equation}
  r_{hq} = \frac{(\zeta_{q} - \zeta_{h})} {(\zeta_{q} + \zeta_{h})},
\end{equation}
and
\begin{equation}
  \zeta_{h}=\frac{e_h}{k_{zh}},
\end{equation}
where R is the final reflection output; $i = 0$ to $7$; $d_1;d_2;d_3;d_4;d_5;d_6;d_7$ and $d_8$ are the thicknesses
of metal, first MoS$_2$ layer, first graphene layer, first ligand layer, sensing
layer, second ligand layer, second graphene layer and second $MoS_2$ layer
respectively; $k_{z0}$ to $k_{z9}$ are the wavevectors of the respective media; $h = 0$ to $9$ and $r_{01}, r_{12}, r_{23}, r_{34}, r_{45}$ are the amplitude
reflectances given by Fresnel formulas of p-polarization for prism-metal,
metal-$MoS_2$, $MoS_2$-graphene, graphene-tryptophan, tryptophan-lipid
intefaces in order on one side of the channel. Again, $r_{56}, r_{67},
r_{78}, r_{89}$ represent reflection coefficients from the lipid-tryptophan,
tryptophan-graphene, graphene-$MoS_2$ and $MoS_2$-air interfaces,
respectively; $e_0;e_1;e_2;e_3;e_4;e_5$ are the dielectric constants of the
prism, metal, $MoS_2$, graphene, tryptophan and sensing layer. $e_4$ and $e_6$
are equal, $e_3$ and $e_7$ are same, $e_2$ and $e_8$ are same, $e_9$ corresponds to air
and
\begin{equation}
   k_{zh} = (\frac{2\times\pi}{\lambda})\times\sqrt(e_h - e_o\times(\sin\theta)^2). 
\end{equation}

\section{Simulation Results}

For simulation purposes, we model the source as a He-Ne laser having 632.8 nm emission
wavelength. This wavelength source has been chosen because of relatively low cost and ease of visible-range operation, producing beams of similar quality in terms of spatial coherence and long coherence length. The incident light is simulated to be TM polarized. To investigate the effect of incidence angle change, the source and the detector are rotated synchronously at small angular steps, keeping the sensor setup stationary. The sensitivity of an SPR sensor can be defined as
\begin{equation}
   S_{\theta} = \frac{d \theta_{SPR}}{d n_s},
\end{equation}
\noindent
where $d_{SPR}$ is the differential change in SPR angle due to $dn_s$,
differential change in sample layer refractive index.

\subsection{Parameter Settings}
In this section,
we discuss the modeling of this basic structure in FDTD simulator, providing parameters used in the simulation and convergence test. The meshing was adjusted to provide convergence with sufficient precision, without making the simulation unnecessarily lengthy. The material layers were simulated as layers with corresponding complex refractive indices in the respective wavelengths. For intermediate arbitrary wavelength, interpolated values are used. We used
data obtained by Palik \cite{palik} to model the layers. The estimated index of refraction for graphene was 3+1.1487i using $n = 3 + iC\lambda/3$, where $\lambda$ is wavelength ($\mu m$) and C is 5.446 $\mu m^{-1}$ \cite{zeng}. The refractive index of MoS$_2$ was estimated to be approximately 5.9+0.83i \cite{refindmos2}. Although bio-layers are not necessarily continuous layers, for simplicity
we approximated the ligand-ligate layers as a single continuous layer. Again, the layers have been modeled as dielectrics with complex refractive
indices obtained from \cite{biorefind}. For our simulations, we applied perfectly matched layer (PML) boundary
conditions along the direction of light incidence and bloch boundary
conditions along the direction perpendicular to it. 
%Plane wave sources were used to inject laterally-uniform electromagnetic
%energy from one side of the source region.

\subsection{Performance analysis}
In Fig.~2(c), we show reflectance against angle of incidence. In Fig.~2(e), $\theta_{SPR}$ is plotted as a function of the sample's index of refraction, we fit a linear least square error curve passing through the dip angles. We determine
the sensitivity from the slope of this line and it was found that for the
proposed structure,  $S_{\theta}$ = $75.1954^{\degree}/RIU$ while for basic Kretschmann, $S_{\theta}$ = $17.9138^{\degree}/RIU$. So, a significant improvement is
obtained by adopting this approach over the conventional metal-dielectric
configuration and the ratio of the sensitivity of the proposed one to basic
one is about 4.2. We also noticed in Fig. 2(d) that maximum power is coupled at the resonance angle which validates the occurrence of SPR coupling at that angle.
\begin{figure}[H]
\centering
\graphicspath{ {Thesis/} }
\includegraphics[height=4in, width = 6.5in]{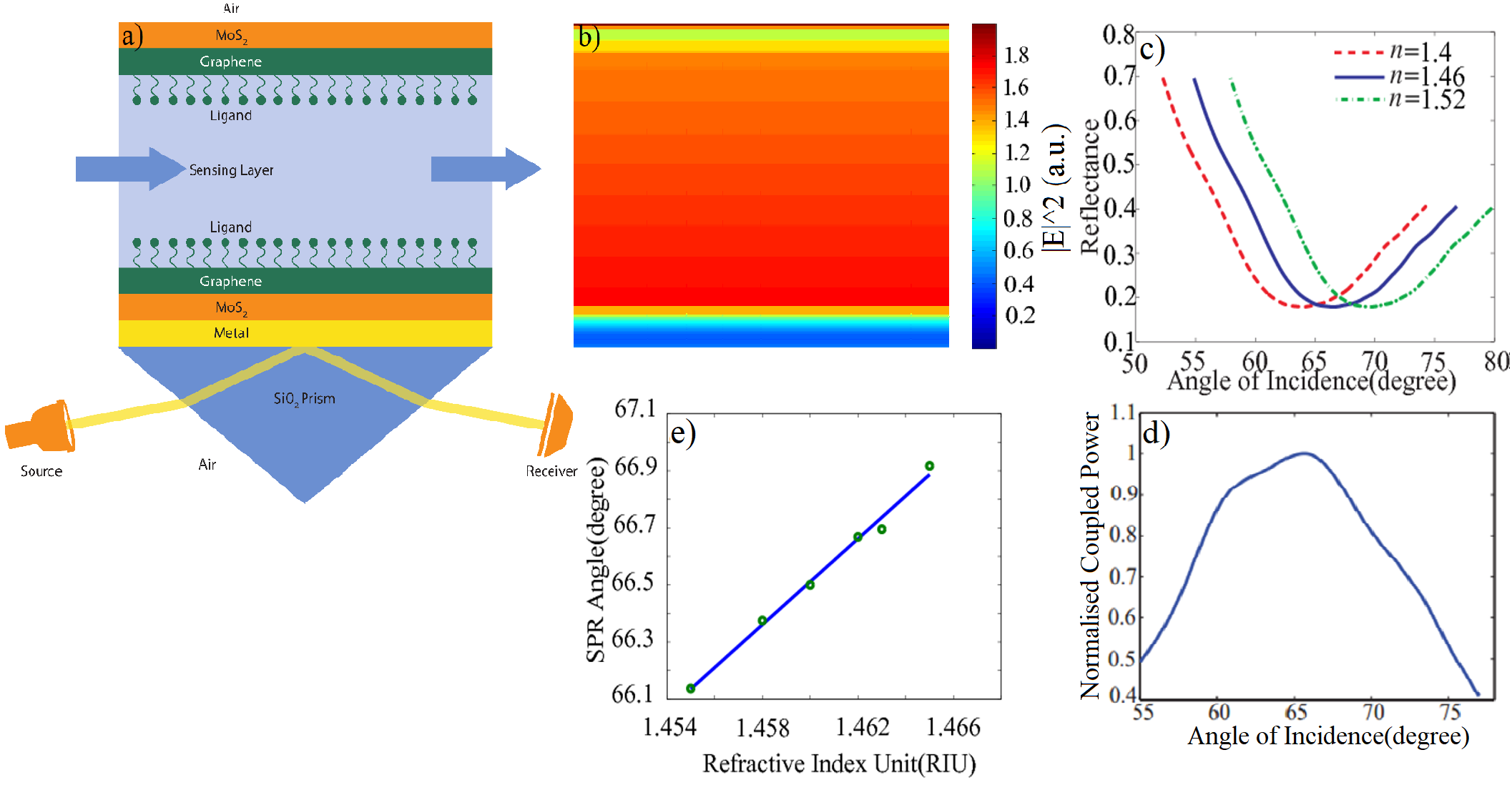} 
\caption{(a) Schematic of the proposed structure, (b) Electric field intensity profile in the proposed
structure for TM incident light at a wavelength of 632.8 nm, (c) Reflected light intensity versus
source incidence angle at a wavelength of 632.8 nm, (d) Normalised coupled power versus angle of incidence at a wavelength of 632.8 nm with the sample of 1.46 refractive index, (e) SPR angle versus sample refractive index
to calculate sensitivity.}
\end{figure}
The electric field profile across the direction of field penetration inside
the sensing layer is shown in Fig. 2(b). As can be seen from here, the
E field penetrates through the channel and reaches the upper ligand layer upto the layer of $MoS_2$. This is what enhances the sensitivity. If the
molecule dimensions are small enough then the evanescent wave can reach through them and go upto the next layers as demonstrated in Fig.~3. As a result, the change in the effective index of the sensing layer is
more after ligand-analyte bonding. This is the physical reason behind
the increased sensitivity of the structure. 
\begin{figure}[H]
\centering
\graphicspath{ {Thesis/} }
\includegraphics[height=2.2in, width = 3.5in]{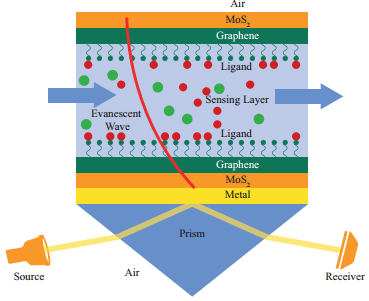} 
\caption{Penetration of evanescent wave through the proposed bilayer structure (here red and green circles represent target analytes and solvent molecules respectively).}
\end{figure}
Though we focus mainly on the angle interrogation technique, our structure provides high sensitivity in wavelength interrogation as well. To characterize the proposed framework in terms of wavelength criterion, we use a broadband source having a spectra between 400 to 700 nm. Also we set the source incidence angle such that at the metal interface the light is incident at $66.54^{\degree}$. Then we measure the reflected light at different wavelengths sweeping the center wavelength and observe the reflectance dip. Fig.~4 shows the reflectance using the sample layer indices 1.4, 1.46
and 1.52. Here we see that the wavelength corresponding to the dip
experiences a red shift as the index rises. Also from here we calculate the sensitivity in terms of the wavelength shift as
\begin{equation}
 S_\lambda = \frac{d\lambda}{dn}, 
\end{equation}
\begin{figure}[H]
\centering
\graphicspath{ {Thesis/} }
\includegraphics[height=2.2in, width = 3in]{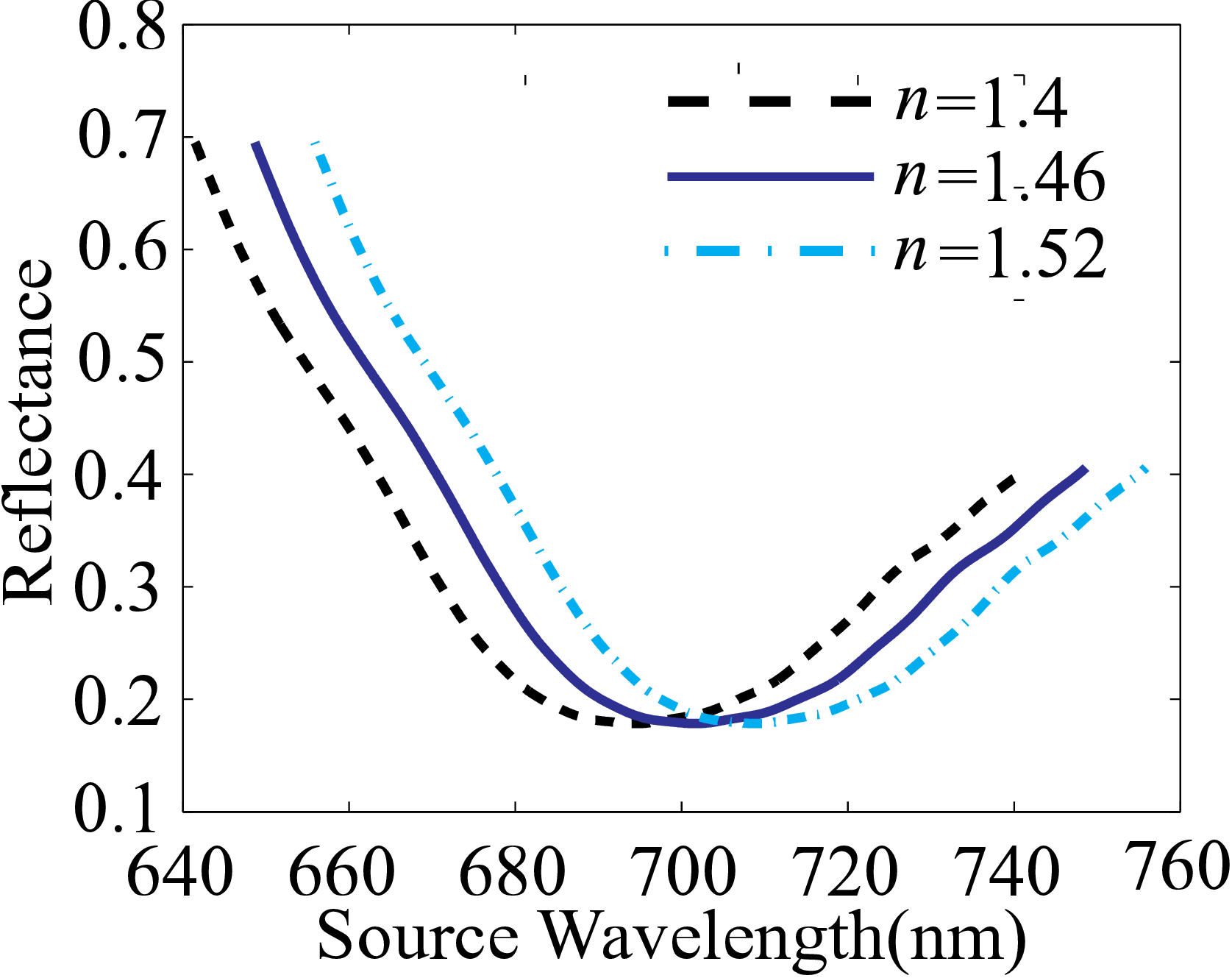} 
\caption{Reflected light intensity versus
source wavelength at an angle of 66.54.}
\end{figure}
where $d\lambda$ is the differential change in SPR wavelength due to
$dn$, differential change in sample layer refractive index. From the
observations, we find with the proposed approach, the sensitivity is,  $S_{\lambda}$ = 116.67 nm/RIU. Again in this regard, the sensitivity of the Kretschmann
structure was simulated and found to be 20 nm/RIU. So, once again we
find that the designed structure gives superior performance over the basic
one.

We analyzed all structures in the process of the development of proposed structures. The results are shown in Figs.~5 and 6. The superiority of our structure is obvious from the graphs. 

\begin{figure}[H]
\centering
\graphicspath{ {Thesis/} }
\includegraphics[height=3.5in, width = 5in]{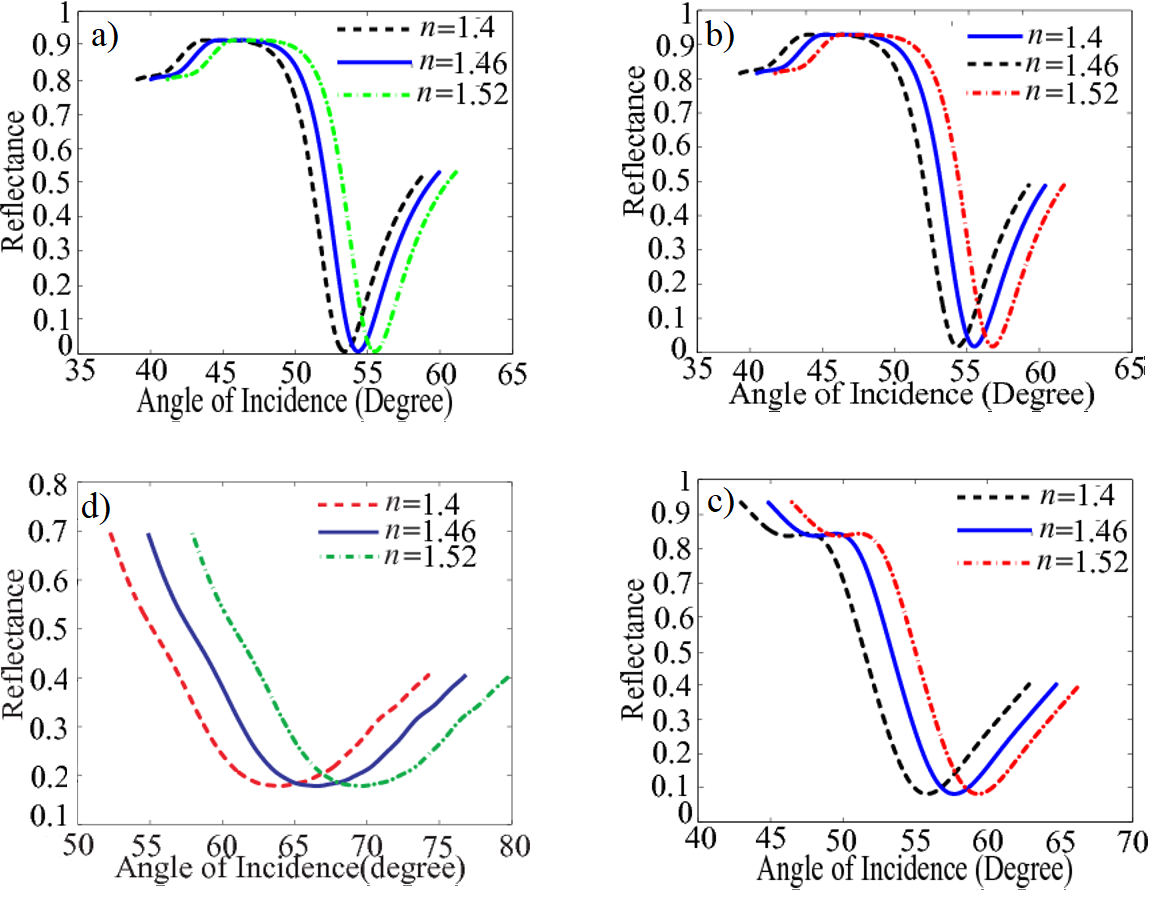} 
\caption{Reflectance versus incidence angle of (a) Basic structure (b) Bilayer structure (c) Bilayer structure with graphene (d) proposed structure.}
\end{figure}
We varied the refractive index from 1.4 to 1.52 and observed the shift in angle of incidence in Fig.~5. Then, we plotted the shift of the angle of incidence with the refractive index variation in Fig.~6 . The slopes of the graphs give us corresponding sensitivity. We find that for basic Kretschmann, $S_{\theta}$ = $17.9138^{\degree}/RIU$. For, Fig.~6(b), we found sensitivity, $S_{\theta}$ = $20.2^{\degree}/RIU$. As we can see, introducing the bilayer aids in higher binding of the
analytes which alters the index by a higher degree. This in turn results in a larger shift in SPR angle which enhances the sensitivity. The sensitivity obtained for the bilayer with graphene case from Fig.~6(c) was, $S_{\theta}$ = $25.02^{\degree}/RIU$ and finally, for the proposed structure, it was $S_{\theta}$ = $75.2^{\degree}/RIU$, as mentioned before. It is clearly noticeable that the sensitivity gradually increased as we went from basic to the proposed structure. To further validate the pertinence of the proposed configuration, various disease conditions have also been simulated \cite{our1} and results show that the structure is able to successfully
detect them to a satisfactory limit.
\begin{figure}[H]
\centering
\graphicspath{ {Thesis/} }
\includegraphics[height=3.5in, width = 5in]{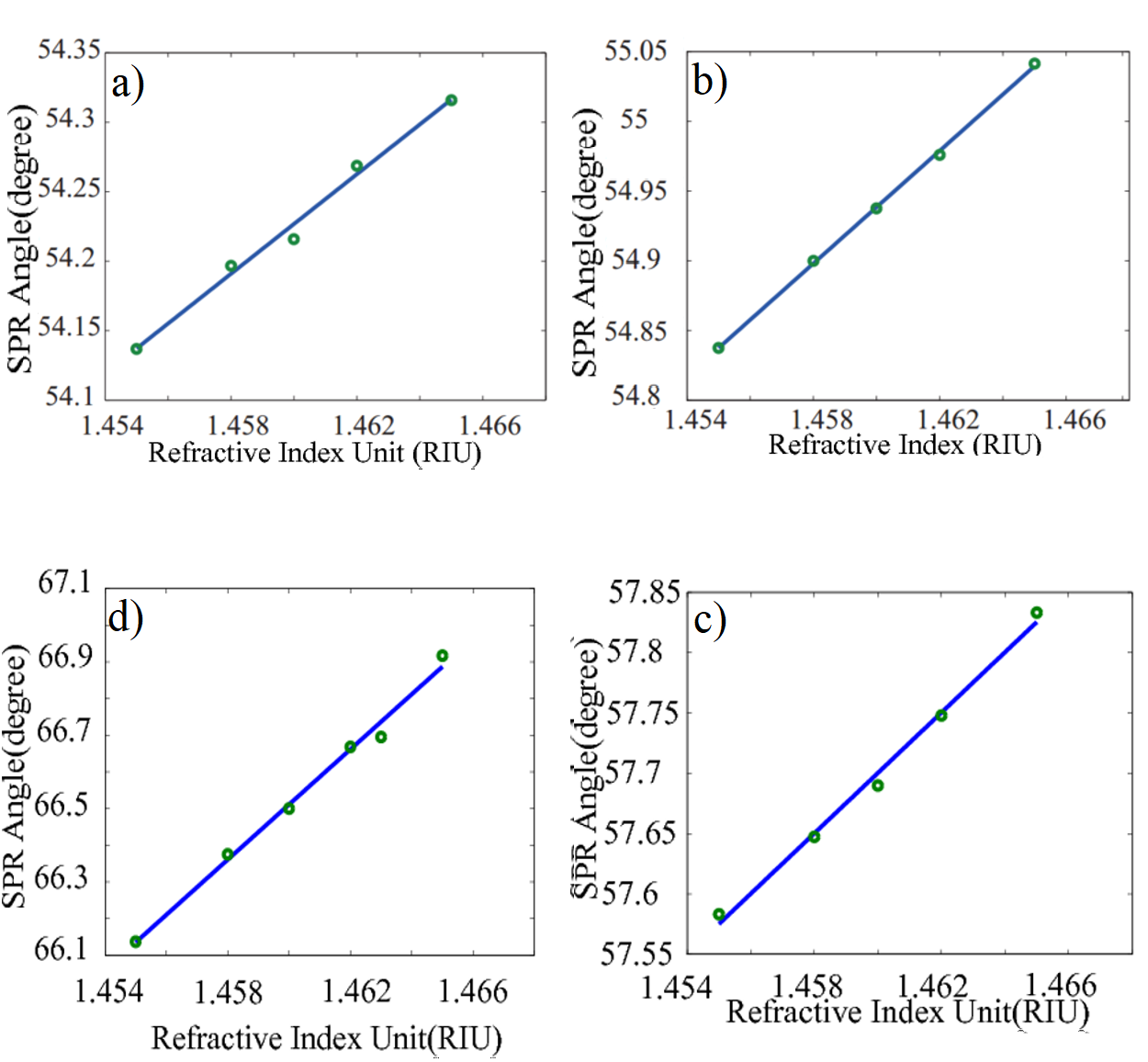} 
\caption{SPR angle versus sample refractive index of (a) basic structure, (b) bilayer structure, (c) bilayer structure with graphene and (d) proposed structure.}
\end{figure}
\subsection{Comparison with other Enhancement Structures}
The proposed structure provides significant enhancement as it has been
described above. In Table 1, we provide the normalized sensitivity of
the proposed structure, as well as of structures of other Refs. If
we compare the performance, we can clearly see the relative improvement provided by the bilayer architecture.
\begin{table}[H]
\centering
\caption{Ratio of sensitivity for proposed and other structures over
basic Kretschmann configuration}
\label{my-label}
\begin{tabular}{ll}
\hline
Structure                                    & $S_{\theta}(normalized) = S/S_{Kretschmann}$ \\ \hline
Proposed Structure                           & 4.2                                          \\

Nanowire Based SPR sensor \cite{nanowire}                  & 1.42                                         \\

Graphene coated SPR sensor\cite{rahman} & 2.2                                         \\

ZnO assisted Graphene-based SPR sensor \cite{kumar} & 1.27                                         \\
Trilayered Metallic Structure \cite{trilayer}       & 1.53                                         \\
Graphene Oxide Multilayer Structure \cite{chung} & 1.03                                         \\

Cr-Ag-ITO SPR Refractive Sensor \cite{Gan}     & 3.90                                         \\
\hline
\end{tabular}
\end{table}
\subsection{Robustness of Sensitivity Against Choice of Ligand-ligate Pair}
We studied the performance of this structure for other lipids using
different suitable ligands and obtained similar sensor characteristics.
Table 2 shows the various S values obtained for the combinations. We
see that the sensitivity enhancement is fairly consistent over all the pairs.
\begin{table}[H]
\centering
\caption{Comparison of Sensitivity among different ligand-analyte
combinations}
\label{my-label}
\begin{tabular}{l|l}
\rowcolor[HTML]{F8FF00} 
Ligand-ligate Pair                           & Sensitivity, S(degree/RIU) \\ \hline
\multicolumn{1}{|l|}{Tryptophan-Phospolipid} & \multicolumn{1}{l|}{75.2}  \\ \hline
\multicolumn{1}{|l|}{BSA-Phospolipid}        & \multicolumn{1}{l|}{74.7}  \\ \hline
\multicolumn{1}{|l|}{BSA-Egg Yolk}           & \multicolumn{1}{l|}{75.03} \\ \hline
\end{tabular}
\end{table}

\section{Conclusion}

Sensitivity enhancement is really important specially for detecting very small changes in the sample layer which
otherwise can easily go unnoticed. If we have a large number of cell samples and only a
portion of them are infected due to attack of a disease, then the corresponding refractive index variation of the sample layer could be quite small. In those cases, to ensure the presence or absence of a certain phenomenon or to detect a disease, the more sensitive sensor
can prove useful whereas the less sensitive ones might fail to detect them properly. Hence with sensitivity enhancement as primary focus, we propose a mirrored
bilayer structure here incorporating layers of graphene and MoS$_2$ and validate our claims of enhanced sensitivity  through FDTD simulations.
Moreover, the structure can provide enhancement for both angular and wavelength interrogations. As a result, the proposed configuration, if implemented properly, can yield promising results as evidenced from the comparison with other state-of-the art sensor designs.

%% References



\begin{thebibliography}{99}

\bibitem{kurihara} K. Kurihara, and K. Suzuki, ``Theoretical Understanding of an Absorption-Based Surface Plasmon Resonance Sensor Based on Kretchmann's Theory,'' Anal. Chem. \textbf{74}, 696--701 (2002). 

\bibitem{ppg1} S. S. Chowdhury, R. Hyder, M. S. B. Hafiz, and M. A. Haque, ``Real Time Robust Heart Rate Estimation from Wrist-type PPG Signals Using Multiple Reference Adaptive Noise Cancellation,'' IEEE Journal of Biomedical and Health Informatics \textbf{22(2)}, 450--459 (2016). 

\bibitem{ppg2} S. S. Chowdhury, M. S. Hasan, and R. Sharmin, ``Robust Heart Rate Estimation from PPG Signals with Intense Motion Artifacts using Cascade of Adaptive Filter and Recurrent Neural Network,'' in \textit{IEEE Region 10 Conference (TENCON),}  India (2019). 


\bibitem{2} R. Liu, Q. Wang, Q. Li, X. Yang, K. Wang, and W. Nie, ``Surface plasmon resonance biosensor for sensitive detection of
microRNA and cancer cell using multiple signal amplification strategy," Biosensors and Bioelectronics \textbf{87}, 433-–438 (2017).

\bibitem{nanowire} K. Byun, S. Yoon, D. Kim and S. Kim, ``Experimental study of sensitivity enhancement in surface plasmon resonance biosensors by use of periodic metallic nanowires,'' Optics Letters \textbf{32}, 1902--1904 (2007).


%\bibitem{5} J. Homola, S. S. Yee, G. Gauglitz, ``Surface plasmon resonance
%sensors: review”, \textit{Sens. Actuators} B 54, 3-15 (1999).

%\bibitem{6} Yuhki Yanase, Takaaki Hiragun, Tetsuji Yanase, Tomoko Kawaguchi, Kaori Ishii,
%Nobutaka Kumazaki, Takayuki Obara, Michihiro Hide, ``Clinical diagnosis of type I allergy by means of SPR imaging with less than a microliter of peripheral blood," \textit{Sensing and Bio-Sensing Research} 2 (2014) 43–48. 

\bibitem{collagen} J.-M.~Friedt, and L.~A.~Francis, ``Combined surface acoustic wave and surface plasmon resonance measurement of collagen and fibrinogen layer physical properties," Sensing and Bio-Sensing Research \textbf{11}, 60--70 (2016).

%\bibitem{8} Ivan Stojanović, Yoeri van Hal, Thomas J.G. van der Velden,
%Richard B.M. Schasfoort, Leon W.M.M. Terstappen, ``Detection of apoptosis in cancer cell lines using Surface Plasmon Resonance imaging," \textit{Sensing and Bio-Sensing Research} 7 (2016) 48–54.

\bibitem{ecoli} C.~Zhou, H.~Zou, M.~Li, C.~Sun, D.~Ren, and Y.~Li, ``Fiber optic surface plasmon resonance sensor for detection of E. coli O157:H7 based on antimicrobial peptides and AgNPs-rGO," Biosensors and Bioelectronics \textbf{117}, 347--353 (2018).

%\bibitem{10} Stefania Torinoa, Laura Conteb, Mario Iodice, Giuseppe Coppola, Ralf D. Prien, ``PDMS membranes as sensing element in optical sensors for gas detection in
%water," \textit{Sensing and Bio-Sensing Research} 16 (2017) 74–78.

%\bibitem{11} Jean-Francois Masson, `` Surface Plasmon Resonance Clinical Biosensors for Medical Diagnostics," \textit{ACS Sens.,} 2017, 2 (1), pp 16--30.

\bibitem{our1} S.~S.~Chowdhury, S.~M.~A.~Uddin, E.~Kabir and A. M. Mahmud Chowdhury, ``Detection of DNA Mutation, Urinary Diseases and Blood Diseases using Surface Plasmon Resonance Biosensors Based on Kretschmann Configuration," in \textit{International Conference on Electrical, Computer and Communication Engineering (ECCE),} Bangladesh (2017).

\bibitem{liedberg} B. Liedberg, C. Nylander, and I. Lundstrom, \quotes{Surface plasmon resonance for gas detection and biosensing}, Sens. Actuators \textbf{4}, 299–-304 (1983).

\bibitem{homola} J. Homola, \quotes{Surface plasmon resonance sensors for detection of chemical and biological species}, Chem. Rev. \textbf{108}, 462–-493 (2008).
\bibitem{otto} A. Otto, \quotes{Excitation of Nonradiative surface plasma waves in silver by
the method of frustrated total reflection}, Z. Physik \textbf{216}, 398--410 (1968).
\bibitem{krets} E. Kretschmann, and Raether, \quotes{Radiative decay of nonradiative surface
plasmons excited by light}, Z. Naturforsch, \textbf{23A}, 2135--2136 (1968).
\bibitem{nylander} C. Nylander, B. Liedberg, and T. Lind, \quotes{Gas detection by means of
surface plasmon resonance}, Sens. Actuators \textbf{3}, 79--88 (1982).

%\bibitem{yuan} B. Yuan, X. Jiang, C. Yao, M. Bao, J. Liu,
%Y. Dou, Y. Xu, Y. He, K. Yang, and Y. Ma, ``Plasmon-enhanced fluorescence imaging with silicon-based silver chips for protein and nucleic acid assay," Analytica Chimica Acta \textit{955} (2017) 98e107.

\bibitem{suvarna} P. Suvarnaphaet and S. Pechprasarn, ``Quantitative Cross-Platform Performance Comparison between Different Detection Mechanisms in Surface Plasmon Sensors for
Voltage Sensing," Sensors \textbf{18}, 31--36 (2018).


%\bibitem{21} Chutiparn Lertvachirapaiboon, Akira Baba, Sanong Ekgasit, Kazunari Shinbo, Keizo Kato, Futao Kaneko, ``Transmission surface plasmon resonance imaging of silver
%nanoprisms enhanced propagating surface plasmon resonance on a metallic grating structure," \textit{Sensors and Actuators B} 249 (2017) 39–43.

\bibitem{lertva} C. Lertvachirapaiboona, A. Babaa, S. Ekgasit, K. Shinbo,
K. Kato, and F. Kaneko, ``Transmission surface plasmon resonance techniques and their
potential biosensor applications," Biosensors and Bioelectronics \textbf{99}, 399--415 (2018).

\bibitem{chiu} N. Chiu, and T. Huang, \quotes{Sensitivity and kinetic analysis of graphene oxide-based surface plasmon resonance biosensors}, Sensors and Actuators B \textbf{197}, 35–-42 (2014).

\bibitem{zeng} S.~Zeng, S.~Hu, J.~Xia, T.~Anderson, X.~Dinh,
X. Meng, P. Coquet, and K. Yong, ``Graphene–-MoS$_2$ hybrid nanostructures enhanced surface plasmon resonance biosensors," Sensors and Actuators B \textbf{207}, 801–-810 (2015).

%\bibitem{25} Jonghwan Kim, Hyungmok Son, David J. Cho, Baisong Geng, Will Regan, Sufei Shi,
%Kwanpyo Kim, Alex Zettl, Yuen-Ron Shen and Feng Wang, ``Electrical Control of Optical Plasmon Resonance with Graphene," \textit{Nano Lett.} 2012, 12, 5598−5602.

%\bibitem{26} Pradeep Kumar Maharana, Rajan Jha, ``Chalcogenide prism and graphene multilayer based surface plasmon resonance affinity biosensor for high performance," \textit{Sensors and Actuators B} 169 (2012) 161–166.

\bibitem{rahman} M. S. Rahman, M. S. Anower, L. B. Bashar, and K. A. Rikta, ``Sensitivity analysis of graphene coated surface plasmon resonance
biosensors for biosensing applications," Sensing and Bio-Sensing Research \textbf{16}, 41–-45 (2017) .

%\bibitem{28} L. Wu, H. S. Chu, W. S. Koh, and E. P. Li, ``Highly sensitive graphene biosensors based on surface plasmon resonance," \textit{OPTICS EXPRESS} 5 July 2010 / Vol. 18, No. 14.

\bibitem{kumar} R. Kumar, A. S. Kushwaha, M. Srivastava, H. Mishra and S. K. Srivastava, ``Enhancement in sensitivity of graphene-based zinc oxide assisted
bimetallic surface plasmon resonance (SPR) biosensor," Applied Physics A \textbf{207}, 124–-235 (2018).

%\bibitem{30} J. B. Haun, T. J. Yoon, H. Lee, and R. Weissleder, \quotes{Magnetic
%nanoparticle biosensors}, Wiley Interdiscip. Rev. Nanomedicine Nanobiotechnology \textbf{2}, 291-–304 (2010).

\bibitem{sun} Y. Sun, K. Liu, Y. Han, Q. Li, S. Fan, and K. Jiang, \quotes{Excitation
of surface plasmon resonance in composite structures based on
single-layer superaligned carbon nanotube films}, J. Phys. Chem.
C \textbf{117}, 23190-–23197 (2013).

\bibitem{jia} Y. Jia, Y. Peng, J. Bai, X. Zhang,Y. Cui, B. Ning, J. Cui and Z. Gao, ``Magnetic Nanoparticle Enhanced Surface PlasmonResonance sensor for estradiol analysis," Sensors and Actuators B \textbf{254}, 629–-635 (2018).

\bibitem{li} G. Li, X. Li, M. Yang, M. M. Chen, L. C. Chen, and X. L. Xiong,
\quotes{A gold nanoparticles enhanced surface plasmon resonance
immunosensor for highly sensitive detection of Ischemia Modified Albumin}, Sensors \textbf{13}, 12794--12803 (2013).

%\bibitem{33} J. Matsui, K. Akamatsu, N. Hara, D. Miyoshi, H. Nawafune, K.
%Tamaki, and N. Sugimoto, \quotes{SPR sensor chip for detection of
%small molecules using molecularly imprinted polymer with
%embedded gold nanoparticles}, Anal. Chem. \textbf{77}, 4282-–4285 (2005).

%\bibitem{34} R.~B.~Pernites, R.~R.~Ponnapati, and R.~C.~Advincula, \quotes{Surface
%plasmon resonance (SPR) detection of theophylline via
%electropolymerized molecularly imprinted polythiophenes},
%Macromolecules \textbf{43}, 9724-–9735 (2010).

\bibitem{du} W. Du, and F. Zhao, \quotes{Surface plasmon resonance based silicon
carbide optical waveguide sensor}, Mater. Lett. \textbf{115}, 92–-95 (2014).

\bibitem{rowe} D. J. Rowe, J. S. Jeong, K. A. Mkhoyan, and U. R. Kortshagen,
\quotes{Phosphorus-Doped silicon nanocrystals exhibiting mid-infrared
localized surface plasmon resonance}, Nano Lett. \textbf{13},
1317-–1322 (2013). 

%\bibitem{38} S. P. Ng, Y. Y. Yip, and C. L. Wu, \quotes{Biosensing with gain-assisted surface plasmon-polariton amplifier: A computational investigation}, Sensors and Actuators B \textbf{210}, 36–-45 (2015). 

\bibitem{trilayer} Z. Wang, Z. Cheng, V. Singh, Z. Zheng, Y. Wang, S. Li, L. Song, and J. Zhu, \quotes{Stable and Sensitive Silver Surface Plasmon Resonance Imaging
Sensor Using Trilayered Metallic Structures}, Anal. Chem. \textbf{86}, 1430--1436 (2014).

\bibitem{artar} A. Artar, A. A. Yanik, and H. Altug, \quotes{Fabry–-Perot nanocavities in multilayered plasmonic crystals for enhanced biosensing}, Applied Physics Letters \textbf{95}, 051105 (2009).

\bibitem{shuwen} S. Zeng, \quotes{Sensitivity improved surface plasmon resonance biosensor for cancer biomarker detection based on 2D perovskite-based metasurfaces }, Plasmonics in Biology and Medicine XVI, Proceedings Volume 10894, (2019).

%\bibitem{42 } S. Kubitschko, J. Spinke, T. Bruckner, S. Pohl, and N. Oranth, \quotes{Sensitivity enhancement of optical immunosensors with nanoparticles}, Anal. Biochem. \textbf{253}, 112-–122 (1997).

\bibitem{chung} K. Chung, A. Rani, J. Lee, J. Kim, Y. Kim, H. Yang, S. O. Kim, D. Kim, and D. H. Kim, \quotes{Systematic Study on the Sensitivity Enhancement in Graphene Plasmonic Sensors Based on Layer-by-Layer Self-Assembled Graphene Oxide Multilayers and Their Reduced Analogues}, ACS Appl. Mater. Interfaces \textbf{7}, 144--151 (2015).


\bibitem{our2} S. M. A. Uddin, S. S. Chowdhury, and E. Kabir, ``A Theoretical Model for Determination of Optimum Metal Thickness in Kretschmann Configuration Based Surface Plasmon Resonance Biosensors", in \textit{International Conference on Electrical, Computer and Communication Engineering (ECCE),} Bangladesh (2017).

\bibitem{maurya} J. B. Maurya, Y. K. Prajapati, and R. Tripathi, \quotes{Effect of Molybdenum Disulfide Layer on Surface Plasmon Resonance Biosensor for the Detection of Bacteria}, Silicon \textbf{10 (2)}, 245--256 (2018).

\bibitem{palik}  E. D. Palik, \textit{Handbook of optical constants of solids}, volume 3.
Academic press, (1998).

\bibitem{refindmos2} A. Castellanos-Gomez, N. Agraït, and G. Rubio-Bollinger, ``Optical identification of atomically thin dichalcogenide crystals ," Appl. Phys. Lett. \textbf{96}, 213116 (2010).

\bibitem{biorefind} T. L. Mcmeekin, M. L. Grooves and N. J. Hipp, \quotes{Refractive Indices of Amino Acids, Proteins, and Related Substances}, Advances in Chemistry \textbf{44}1, 54--66 (1964).

%\bibitem{Gramultilayer} K. Chung, A. Rani, J. Lee, J. Kim, Y. Kim, H. Yang, S. O. Kim, D. Kim and D. H. Kim, ``Systematic Study on the Sensitivity Enhancement in Graphene Plasmonic Sensors Based on Layer-by-Layer Self-Assembled Graphene Oxide Multilayers and Their Reduced Analogues,'' ACS Appl. Mater. Interfaces \textbf{7}, 144--151 (2015).

\bibitem{Gan} S. M. Gan, P. S. Menon, N. R. Mohamad, N. A. Jamil, and B. Y. Majlis, ``FDTD simulation of Kretschmann based Cr-Ag-ITO SPR for refractive index sensor," \textit{Materials Today: Proceedings} Volume 7, Part 2, 2019, Pp. 668-674.

%\bibitem{49} O. Zhernovaya, O. Sydoruk, V. Tuchin1, and A. Douplik,  \quotes{The refractive index of human hemoglobin in the visible range}, Phys. Med. Biol. \textbf{56}, 4013-–4021 (2011).

%\bibitem{50} H. Ghandoor, H. E. Yasser, A. F. El-Sherif, and R. M. Elghwas, \quotes{Determination of Glucose Concentration Using Optical Coherence Tomography}, in \textit{13\textsuperscript{th} International Conference on Aerospace Sciences and Aviation Technology}.

\end{thebibliography}
\end{document}